\documentstyle[aps,prl,twocolumn,epsf,floats]{revtex}
\begin{document}
\wideabs{
\title{Two-Dimensional Electron-Hole Systems in a Strong Magnetic Field}
\author{
   John J. Quinn$^1$,
   Arkadiusz W\'ojs$^{1,2}$, 
   Izabela Szlufarska$^{1,2}$, 
   and Kyung-Soo Yi$^{1,3}$}
\address{
   $^1$Department of Physics, 
       University of Tennessee, Knoxville, Tennessee 37996, USA \\
   $^2$Institute of Physics, 
       Wroclaw University of Technology, Wroclaw 50-370, Poland \\
   $^3$Physics Department, 
       Pusan National University, Pusan 609-735, Korea}

\maketitle

\begin{abstract}
   Two-dimensional systems containing $N_e$ electrons and $N_h$ holes
   ($N_e>N_h$) strongly correlated through Coulomb interactions in the
   presence of a large magnetic field are studied by exact numerical
   diagonalization.
   Low lying states are found to contain neutral ($X^0$) and negatively 
   charged ($X^-$) excitons and higher charged exciton complexes ($X_k^-$, 
   a bound state of $k$ neutral excitons and an electron).
   Representing these states in terms of angular momenta and binding 
   energies of the different exciton complexes, and the pseudopotentials
   describing their interactions with electrons and with one another,
   permits numerical studies of systems that are too large to investigate
   in terms of individual electrons and holes.
   Laughlin incompressible ground states of such a multi-component plasma
   are found.
   A generalized composite Fermion picture based on Laughlin type
   correlations is proposed.
   It is shown to correctly predict the lowest band of angular momentum
   multiplets for different charge configurations of the system for any 
   value of the magnetic field.
\end{abstract}
}

\section{Introduction}
In two-dimensional electron-hole systems in the presence of a strong 
magnetic field, neutral excitons $X^0$ and spin-polarized charged 
excitonic ions $X_k^-$ ($X_k^-$ consists of $k$ neutral excitons bound 
to an electron) can occur\cite{shields,wojs1,palacios,wojs2}.
The complexes $X_k^-$ should be distinguished from spin-unpolarized 
ones (e.g. spin-singlet biexciton or charged exciton) that are found
at lower magnetic fields\cite{kheng} but unbind at very high fields
as predicted by hidden symmetry arguments\cite{lerner}.
The excitonic ions $X_k^-$ are long lived Fermions whose energy spectra
display Landau level structure\cite{wojs1,palacios,wojs2}.
In this work we investigate, by exact numerical diagonalization within
the lowest Landau level, small systems containing $N_e$ electrons and 
$N_h$ holes ($N_e>N_h$) confined to the surface of a Haldane sphere
\cite{haldane1,wu}.
For $N_h=1$ these systems serve as simple guides to understanding
photoluminescence\cite{shields,wojs1,palacios,kheng,macdonald,chen,%
rashba,apalkov}.
For larger values of $N_h$ it is possible to form a multi-component 
plasma containing electrons and $X_k^-$ complexes\cite{wojs2}.
We propose a model\cite{halperin} for determining the incompressible 
quantum fluid states\cite{laughlin} of such plasmas, and confirm the 
validity of the model by numerical calculations.
In addition, we introduce a new generalized composite Fermion (CF) 
picture\cite{jain} for the multi-component plasma and use it to predict 
the low lying bands of angular momentum multiplets for any value of the 
magnetic field.

The single particle states of an electron confined to a spherical surface 
of radius $R$ containing at its center a magnetic monopole of strength
$2S\phi_0$, where $\phi_0=hc/e$ is the flux quantum and $2S$ is an integer,
are denoted by $\left|S,l,m\right>$ and are called monopole harmonics
\cite{haldane1,wu}.
They are eigenstates of $\hat{l}^2$, the square of the angular momentum
operator, with an eigenvalue $\hbar^2l(l+1)$, and of $\hat{l}_z$, the
$z$ component of the angular momentum, with an eigenvalue $\hbar m$.
The energy eigenvalue is given by $(\hbar\omega_c/2S)[l(l+1)-S^2]$, 
where $\hbar\omega_c$ is the cyclotron energy.
The $(2l+1)$-fold degenerate Landau levels (or angular momentum shells)
are labelled by $n=l-S=0$, 1, 2, \dots

\section{Four Electron--Two Hole System}

\begin{figure}[t]
\epsfxsize=3.35in
\epsffile{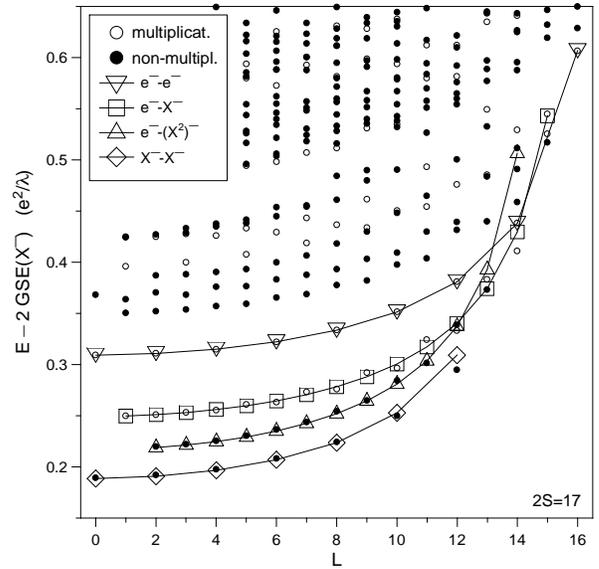}
\caption{
   Energy spectrum of four electrons and two holes at $2S=17$.
   Open circles -- multiplicative states;
   solid circles  -- non-multiplicative states;
   triangles, squares, and diamonds -- approximate pseudopotentials.}
\label{fig1}
\end{figure}

In Fig.~\ref{fig1} we display the energy spectrum obtained by numerical
diagonalization of the Coulomb interaction of a system of four electrons 
and two holes at $2S=17$.
The states marked by open and solid circles are multiplicative
\cite{lerner} (containing one or more decoupled $X^0$'s) and 
non-multiplicative states, respectively.
For $L<12$ there are four rather well defined low lying bands.
Two of them begin at $L=0$.
The lower of these consists of two $X^-$ ions interacting through
a pseudopotential $V_{X^-X^-}(L)$.
The upper band consists of states containing two decoupled $X^0$'s
plus two electrons interacting through $V_{e^-e^-}(L)$.
The band that begins at $L=1$ consists of one $X^0$ plus an $X^-$
and an electron interacting through $V_{e^-X^-}(L)$, while
the band which starts at $L=2$ consists of an $X_2^-$ interacting 
with a free electron.

Knowing that the angular momentum of an electron is $l_{e^-}=S$, 
we can see that $l_{X_k^-}=S-k$, and that decoupled excitons do not 
carry angular momentum ($l_{X^0}=0$).
For a pair of identical Fermions of angular momentum $l$ the allowed
values of the pair angular momentum are $L=2l-j$, where $j$ is an
odd integer.
For a pair of distinguishable particles with angular momenta $l_A$ and 
$l_B$, the total angular momentum satisfies $|l_A-l_B|\le L\le l_A+l_B$.
The states containing two free electrons and two decoupled neutral 
excitons fit exactly the pseudopotential for a pair of electrons at
$2S=17$; the maximum pair angular momentum is $L^{\rm MAX}=16$ as expected.
The states containing two $X^-$'s terminate at $L=12$.
Since the $X^-$'s are Fermions, one would have expected a state at 
$L^{\rm MAX}=2l_{X^-}-1=14$.
This state is missing in Fig.~\ref{fig1}.
By studying two $X^-$ states for low values of $S$, we surmise that the 
state with $L=L^{\rm MAX}$ does not occur because of the finite size 
of the $X^-$.
Large pair angular momentum corresponds to the small average separation,
and two $X^-$'s in the state with $L^{\rm MAX}$ would be too close to 
one another for the bound $X^-$'s to remain stable.
We can think of this as a ``hard core'' repulsion for $L=L^{\rm MAX}$.
Effectively, the corresponding pseudopotential parameter, $V_{X^-X^-}
(L^{\rm MAX})$ is infinite.
In a similar way, $V_{e^-X^-}(L)$ is effectively infinite for 
$L=L^{\rm MAX}=16$, and $V_{e^-X_2^-}(L)$ is infinite for 
$L=L^{\rm MAX}=15$.

Once the maximum allowed angular momenta for all four pairings $AB$ are 
established, all four bands in Fig.~\ref{fig1} can be roughly approximated 
by the pseudopotentials of a pair of electrons (point charges) with angular 
momenta $l_A$ and $l_B$, shifted by the binding energies of appropriate 
composite particles.
For example, the $X^-$--$X^-$ band is approximated by the $e^-$--$e^-$
pseudopotential for $l=l_{X^-}=S-1$ plus twice the $X^-$ energy.
The agreement is demonstrated in Fig.~\ref{fig1}, where the squares, 
diamonds, and two kinds of triangles approximate the four bands in 
the four-electron--two-hole spectrum.
The fit of the diamonds to the actual $X^-$--$X^-$ spectrum is quite
good for $L<12$.
The fit of the $e^-$--$X^-$ squares to the open circle multiplicative 
states is reasonably good for $L<14$, and the $e^-$--$X_2^-$ triangles
fit their solid circle non-multiplicative states rather well for $L<13$.
At sufficiently large separation (low $L$), the repulsion between 
ions is weaker than their binding and the bands for distinct charge 
configurations do not overlap.

There are two important differences between the pseudopotentials 
$V_{AB}(L)$ involving composite particles and those involving point
particles.
The main difference is the hard core discussed above.
If we define the relative angular momentum ${\cal R}=l_A+l_B-L$ for
a pair of particles with angular momentum $l_A$ and $l_B$, then the
minimum allowed relative angular momentum (which avoids the hard core)
is found to be given by
\begin{equation}
   {\cal R}_{AB}^{\rm min}=2\min(k_A,k_B)+1,
\label{eq1}
\end{equation}
where $A=X_{k_A}^-$ and $B=X_{k_B}^-$.
The other difference involves polarization of the composite particle.
A dipole moment is induced on the composite particle by the electric 
field of the charged particles with which it is interacting.
By associating an ``ionic polarizability'' with the excitonic ion 
$X_k^-$, the polarization contribution to the pseudopotential can 
easily be estimated.
When a number of charges interact with a given composite particle, 
the polarization effect is reduced from that caused by a single charge, 
because the total electric field at the position of the excitonic ion
is the vector sum of contributions from all the other charges, and there
is usually some cancellation.
We will ignore this effect in the present work and simply use the
pseudopotentials $V_{AB}(L)$ obtained from Fig.~\ref{fig1} to describe
the effective interaction.

\section{Eight Electron--Two Hole System}

As an illustration, we first present the results of exact numerical
diagonalization performed on the ten particle system ($8e^-$ and $2h^+$).
We expect low lying bands of states containing the following combinations 
of complexes: (i) $4e^-+2X^-$, (ii) $5e^-+X_2^-$, (iii) $5e^-+X^-+X^0$, 
and (iv) $6e^-+2X^0$.
The total binding energies of these configurations are:
$\varepsilon_{\rm i}=2\varepsilon_0+2\varepsilon_1$, 
$\varepsilon_{\rm ii}=2\varepsilon_0+\varepsilon_1+\varepsilon_2$,
$\varepsilon_{\rm iii}=2\varepsilon_0+\varepsilon_1$, and 
$\varepsilon_{\rm iv}=2\varepsilon_0$.
Here $\varepsilon_0$ is the binding energy of an $X^0$, $\varepsilon_1$
is the binding energy of an $X^0$ to an electron to form an $X^-$, and 
$\varepsilon_k$ is the binding energy of an $X^0$ to an $X_{k-1}^-$
to form an $X_k^-$.
Some estimates of these binding energies (in magnetic units $e^2/\lambda$
where $\lambda$ is the magnetic length) as a function of $2S$ are given 
in Tab.~\ref{tab1}.
\begin{table}
\caption{
   Binding energies $\varepsilon_0$, $\varepsilon_1$, $\varepsilon_2$, 
   and $\varepsilon_3$ of $X^0$, $X^-$, $X_2^-$, and $X_3^-$, 
   respectively, in the units of $e^2/\lambda$.}
\begin{tabular}{rcccc}
 $2S$ & $\varepsilon_0$ & $\varepsilon_1$ 
      & $\varepsilon_2$ & $\varepsilon_3$ \\ \hline
  10 & 1.3295043 & 0.0728357 & 0.0411069 & 0.0252268 \\ 
  15 & 1.3045679 & 0.0677108 & 0.0395282 & 0.0262927 \\ 
  20 & 1.2919313 & 0.0647886 & 0.0381324 & 0.0260328
\end{tabular}
\label{tab1}
\end{table}
Clearly, $\varepsilon_0>\varepsilon_1>\varepsilon_2>\varepsilon_3$.
The total energy depends upon not only the total binding energy, but the 
interactions between all the charged complexes in the system as well. 
All groupings (i)--(iv) contain an equal number of $N=N_e-N_h$ singly 
charged complexes.
However, both angular momenta of involved complexes and the relevant 
hard cores are different.
Which of the groupings has a state with the lowest total repulsion and 
binding energy, i.e. the absolute (possibly incompressible) ground state 
of the electron-hole system, depends on $2S$.
It follows from the mapping between electron-hole and spin-unpolarized
electron systems\cite{lerner} that the multiplicative state $Ne^-+N_hX^0$ 
in which all holes are bound into decoupled $X^0$'s is the absolute ground 
state (only) at the values of $2S$ corresponding to the filling factor 
$\nu=1-1/m=2/3$, 4/5, \dots\ of $N$ (excess) electrons.
At other values of $2S$, a non-multiplicative state containing an $X^-$ 
is likely to have lower energy.

\begin{figure}[t]
\epsfxsize=3.35in
\epsffile{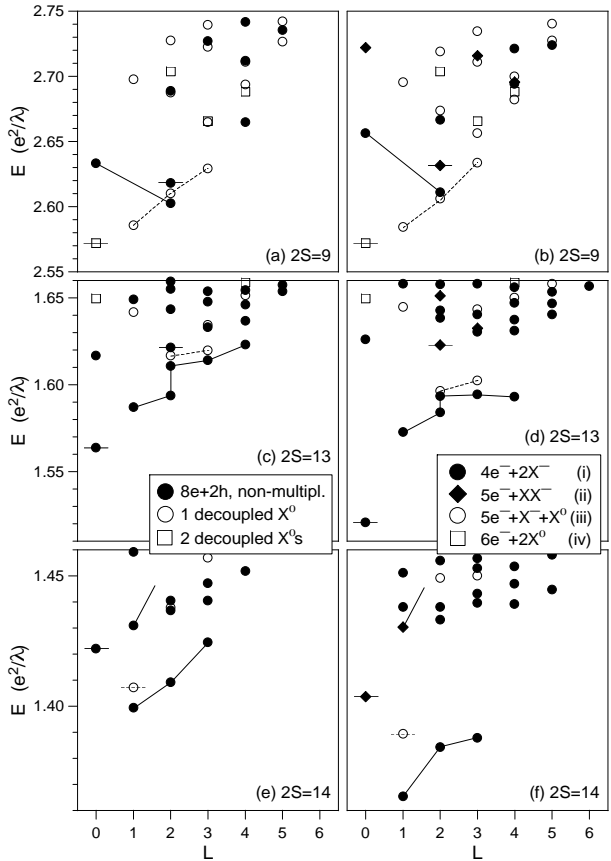}
\caption{
   Left: low energy spectra of the $8e+2h$ system on a Haldane sphere 
   at $2S=9$ (a), $2S=13$ (c), and $2S=14$ (e).
   Right: approximate spectra calculated for all possible groupings 
   containing excitons (charged composite particles interacting through 
   pseudopotentials as in Fig.~\protect\ref{fig1}).
   Lines connect corresponding states in left and right frames.}
\label{fig2}
\end{figure}

In Fig.~\ref{fig2}, we show the low energy spectra of the $8e+2h$ system 
at $2S=9$ (a), $2S=13$ (c), and $2S=14$ (e).
Filled circles mark the non-multiplicative states, and the open circles 
and squares mark the multiplicative states with one and two decoupled
excitons, respectively.
In frames (b), (d) and (f) we plot the low energy spectra of different
charge complexes interacting through appropriate pseudopotentials (see 
Fig.~\ref{fig1}), corresponding to four possible groupings (i)--(iv).
As marked with lines, by comparing left and right frames, we can identify 
low lying states of type (i)--(iv) in the electron-hole spectra.

The fitting of energies in left and right frames at $2S=13$ and 14 is 
noticeably worse than at $2S=9$.
It is also much worse than almost a perfect fit obtained for the three 
charge system ($6e+3h$ vs. $3X^-$, $e^-+X^-+X_2^-$, etc.)\cite{wojs2}.
This is almost certainly due to treating the polarization effect of 
the six charged particle system improperly by using the pseudopotential 
obtained from the two charged particle system (Fig.~\ref{fig1}).
A better fit is obtained by ignoring the polarization effect, and only
including the hard core effect on the pseudopotentials of a pair of
point charges with angular momentum $l_A$ and $l_B$.

\section{Larger Systems}

It is unlikely that a system containing a large number of different
species (e.g. $e^-$, $X^-$, $X_2^-$, etc.) will form the absolute
ground state of the electron-hole system.
However, different charge configurations can form low lying excited 
bands.
An interesting example is the $12e+6h$ system at $2S=17$.
The $6X^-$ grouping (v) has the maximum total binding energy 
$\varepsilon_{\rm v}=6\varepsilon_0+6\varepsilon_1$.
Other expected low lying bands correspond to the following groupings:
(vi) $e^-+5X^-+X^0$ with $\varepsilon_{\rm vi}=6\varepsilon_0+5
\varepsilon_1$ and (vii) $e^-+4X^-+X_2^-$ with $\varepsilon_{\rm vii}
=6\varepsilon_0+5\varepsilon_1+\varepsilon_2$.

\begin{figure}[t]
\epsfxsize=3.35in
\epsffile{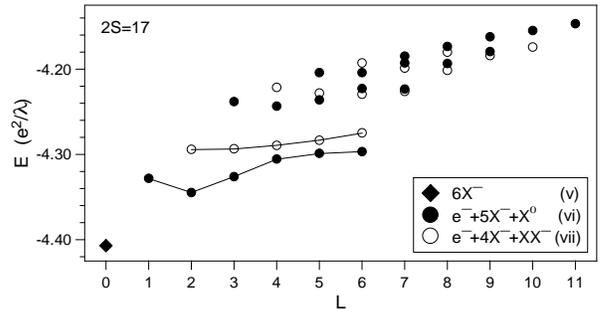}
\caption{
   Low energy spectra of different charge configurations of 
   the $12e+6h$ system on a Haldane sphere at $2S=17$: 
   $6X^-$ (diamonds), $e^-+5X^-+X^0$ (filled circles), 
   and $e^-+4X^-+X_2^-$ (open circles).}
\label{fig3}
\end{figure}

Although we are unable to perform an exact diagonalization for the
$12e+6h$ system in terms individual electrons and holes, we can use
appropriate pseudopotentials and binding energies of groupings (v)--(vii)
to obtain the low lying states in the spectrum.
The results are presented in Fig.~\ref{fig3}.
There is only one $6X^-$ state (the $L=0$ Laughlin $\nu_{X^-}=1/3$ 
state\cite{wojs2}) and two bands of states in each of groupings 
(vi) and (vii).
A gap of 0.0626~$e^2/\lambda$ separates the $L=0$ ground state from 
the lowest excited state.

\begin{figure}[t]
\epsfxsize=3.35in
\epsffile{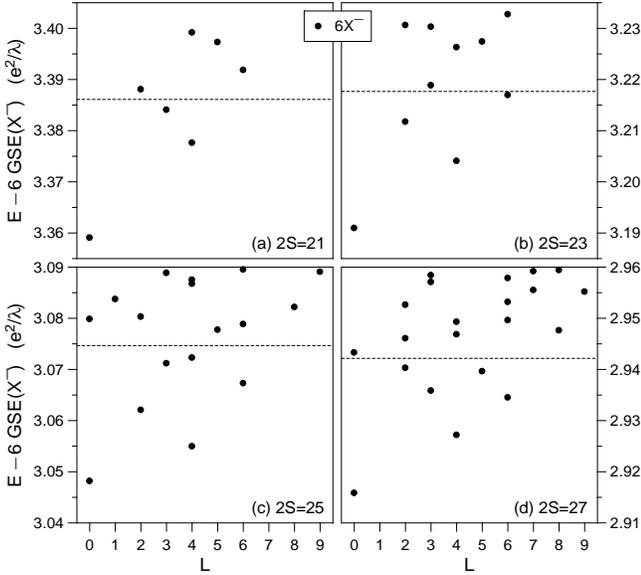}
\caption{
   Low energy spectra of the $6X^-$ charge configuration of the 
   $12e+6h$ system on a Haldane sphere at $2S=21$, 23, 25, and 27.
   Dashed lines -- estimated lower bounds of higher bands.}
\label{fig4}
\end{figure}

In Fig.~\ref{fig4} we present the spectra of the $6X^-$ charge 
configurations for $2S=21$, 23, 25, and 27.
The dashed lines are obtained by adding to the ground state energy 
the binding energy difference appropriate for the next lowest charge
configuration; no states other than the plotted six $X^-$ states are 
expected below these lines.
The $L=0$ ground states observed at different $2S$ correspond to 
$\nu_{X^-}=2/7$, 2/9, 6/29, and 1/5.
The $\nu_{X^-}=1/5$ state is a Laughlin state and $\nu_{X^-}=2/7$ 
and 2/9 are states in Jain sequence.
The $\nu_{X^-}=6/29$ state is a CF hierarchy state\cite{haldane1} 
corresponding to two quasiparticles (QP's) of the $\nu_{X^-}=1/5$ state 
forming a $\nu_{\rm QP}=1/5$ state at the next level of the CF hierarchy.
Without knowing the nature of the QP-QP interaction vs. pair angular 
momentum $L$, there is no guarantee that the CF hierarchy picture 
(which assumes the validity of the mean field approximation) is valid.
Fig.~\ref{fig4}c seems to indicate that it is, since the $L=0$ state
has a lower energy than the other two states at $L=0$ and 4, predicted 
for two QP's each with $l_{\em QP}=5/2$.
Our study of the pseudopotential of QP's in the Laughlin $\nu=1/5$ state 
at $\nu_{\rm QP}=1/5$ very strongly suggests that it behaves like the 
Coulomb pseudopotential, so that the MFCF picture should work.

\section{Generalized Laughlin Wavefunction}

It is known that if the pseudopotential $V({\cal R})$ decreases quickly 
with increasing ${\cal R}$, the low lying multiplets avoid (strongly 
repulsive) pair states with one or more of the smallest values of 
${\cal R}$\cite{wojs3}.
For the (one-component) electron gas on a plane, avoiding pair states 
with ${\cal R}<m$ is achieved with the factor $\prod_{i<j}(x_i-x_j)^m$
in the Laughlin $\nu=1/m$ wavefunction.
For a system containing a number of distinguishable types of Fermions 
interacting through Coulomb-like pseudopotentials, the appropriate
generalization of the Laughlin wavefunction will contain a factor 
$\prod(x^{(a)}_i-x^{(b)}_j)^{m_{ab}}$, where $x^{(a)}_i$ is the 
complex coordinate for the position of $i$th particle of type $a$, 
and the product is taken over all pairs.
For each type of particle one power of $(x^{(a)}_i-x^{(a)}_j)$ results
from the antisymmetrization required for indistinguishable Fermions and
the other factors describe Jastrow type correlations between the 
interacting particles.
Such a wavefunction guarantees that ${\cal R}_{ab}\ge m_{ab}$, for all 
pairings of various types of particles, thereby avoiding large pair 
repulsion\cite{halperin,haldane2}.
Fermi statistics of particles of each type requires that all $m_{aa}$ 
are odd, and the hard cores defined by Eq.~(\ref{eq1}) require that 
$m_{ab}\ge{\cal R}_{ab}^{\rm min}$ for all pairs.

\section{Generalized Composite Fermion Picture}

In order to understand the numerical results obtained in the spherical 
geometry (Figs.~\ref{fig2} and \ref{fig3}), it is useful to introduce 
a generalized CF picture by attaching to each particle fictitious flux 
tubes carrying an integral number of flux quanta $\phi_0$.
In the multi-component system, each $a$-particle carries flux $(m_{aa}-1)
\phi_0$ that couples only to charges on all other $a$-particles and 
fluxes $m_{ab}\phi_0$ that couple only to charges on all $b$-particles,
where $a$ and $b$ are any of the types of Fermions.
The effective monopole strength\cite{chen,jain,wojs3,sitko} seen by 
a CF of type $a$ (CF-$a$) is
\begin{equation}
   2S_a^*=2S-\sum_b(m_{ab}-\delta_{ab})(N_b-\delta_{ab})
\label{eq2}
\end{equation}
For different multi-component systems we expect generalized Laughlin 
incompressible states (for two components denoted as $[m_{aa},m_{bb},
m_{ab}]$) when all the hard core pseudopotentials are avoided and CF's 
of each kind fill completely an integral number of their CF shells 
(e.g. $N_a=2l_a^*+1$ for the lowest shell).
In other cases, the low lying multiplets are expected to contain different 
kinds of quasiparticles (QP-$a$, QP-$b$, \dots) or quasiholes (QH-$a$, 
QH-$b$, \dots) in neighboring filled shells.

Our multi-component CF picture can be applied to the system 
of excitonic ions, where the CF angular momenta are given by 
$l_{X_k^-}^*=|S_{X_k^-}^*|-k$.
As an example, let us first analyze the low lying $8e+2h$ states 
in Fig.~\ref{fig2}.
At $2S=9$, for $m_{e^-e^-}=m_{X^-X^-}=3$ and $m_{e^-X^-}=1$ we predict 
the following low lying multiplets in each grouping:
(i) $2S_{e^-}^*=1$ and $2S_{X^-}^*=3$ gives $l_{e^-}^*=l_{X^-}^*=1/2$.
Two CF-$X^-$'s fill their lowest shell ($L_{X^-}=0$) and we have two 
QP-$e^-$'s in their first excited shell, each with angular momentum 
$l_{e^-}^*+1=3/2$ ($L_{e^-}=0$ and 2).
Addition of $L_{e^-}$ and $L_{X^-}$ gives total angular momenta 
$L=0$ and 2.
We interpret these states as those of two QP-$e$'s in the incompressible 
[331] state.
Similarly, for other groupings we obtain:
(ii) $L=2$;
(iii) $L=1$, 2, and 3; and
(iv) $L=0$ ($\nu=2/3$ state of six electrons).

At $2S=13$ and 14 we set $m_{e^-e^-}=m_{X^-X^-}=3$ and $m_{e^-X^-}=2$ 
and obtain the following predictions.
First, at $2S=13$:
(i) The ground state is the incompressible [332] state at $L=0$;
the first excited band should therefore contain states with one QP-QH 
pair of either kind.
For the $e^-$ excitations, the QP-$e^-$ and QH-$e^-$ angular momenta 
are $l_{e^-}^*=3/2$ and $l_{e^-}^*+1=5/2$, respectively, and the allowed
pair states have $L_{e^-}=1$, 2, 3, and 4.
However, the $L=1$ state has to be discarded, as it is known to have high 
energy in the one-component (four electron) spectrum\cite{sitko}.
For the $X^-$ excitations, we have $l_{X^-}^*=1/2$ and pair states can
have $L_{X^-}=1$ or 2.
The first excited band is therefore expected to contain multiplets at
$L=1$, $2^2$, 3, and 4.
The low lying multiplets for other groupings are expected at:
(ii) $L=2$ and 3;
(iii) $2S_{X_2^-}^*=3$ gives no bound $X_2^-$ state; setting 
$m_{e^-X^-}=1$ we obtain $L=2$; and 
(iv) $L=0$, 2, and 4.
Finally, at $2S=14$ we obtain:
(i) $L=1$, 2, and 3;
(ii) incompressible [3*2] state at $L=0$ ($m_{X^-X^-}$ is irrelevant 
for one $X^-$) and the first excited band at $L=1$, 2, 3, 4, and 5;
(iii) $L=1$; and
(iv) $L=3$.

For the $12e+6h$ spectrum in Fig.~\ref{fig3} the following CF predictions
are obtained:
(v) For $m_{X^-X^-}=3$ we obtain the Laughlin $\nu=1/3$ state with $L=0$.
Because of the hard core of $V_{X^-X^-}$, this is the only state of
this grouping.
(vi) We set $m_{X^-X^-}=3$ and $m_{e^-X^-}=1$, 2, and 3.
For $m_{e^-X^-}=1$ we obtain $L=1$, 2, $3^2$, $4^2$, $5^3$, $6^3$, $7^3$, 
$8^2$, $9^2$, 10, and 11.
For $m_{e^-X^-}=2$ we obtain $L=1$, 2, 3, 4, 5, and 6.
For $m_{e^-X^-}=3$ we obtain $L=1$.
(vii) We set $m_{X^-X^-}=3$, $m_{e^-X_2^-}=1$, $m_{X^-X_2^-}=3$, and 
$m_{e^-X^-}=1$, 2, or 3.
For $m_{e^-X^-}=1$ we obtain $L=2$, 3, $4^2$, $5^2$, $6^3$, $7^2$, $8^2$, 
9, and 10.
For $m_{e^-X^-}=2$ we obtain $L=2$, 3, 4, 5, and 6.
For $m_{e^-X^-}=3$ we obtain $L=2$.
In groupings (vi) and (vii), the sets of multiplets obtained for higher 
values of $m_{e^-X^-}$ are subsets of the sets obtained for lower values, 
and we would expect them to form lower energy bands since they avoid 
additional small values of ${\cal R}_{e^-X^-}$.
However, note that the (vi) and (vii) states predicted for $m_{e^-X^-}=3$ 
(at $L=1$ and 2, respectively) do not form separate bands in 
Fig.~\ref{fig3}.
This is because the $V_{e^-X^-}$ pseudopotential increases more slowly than 
linearly as a function of $L(L+1)$ in the vicinity of ${\cal R}_{e^-X^-}=3$.
In such case the CF picture fails\cite{wojs3}.

The agreement of our CF predictions with the data in Figs.~\ref{fig2}
and \ref{fig3} (marked with lines) is really quite remarkable and 
strongly indicates that our multi-component CF picture is correct.
We were indeed able to confirm predicted Jastrow type correlations in 
the low lying states by calculating their coefficients of fractional 
parentage\cite{wojs3,shalit}.
We have also verified the CF predictions for other systems that we were 
able to treat numerically.
If exponents $m_{ab}$ are chosen correctly, the CF picture works 
well in all cases.

\section{Special Case: Many Electron--One Hole Systems}

In an investigation of photoluminescence, the eigenstates of a system
containing up to $N_e=7$ electrons and a single hole have been studied 
as a function of $d$, the separation between the surfaces on which 
electrons and the hole are confined\cite{chen,rashba,apalkov}.
For $d$ larger than a few magnetic lengths $\lambda$, the low energy 
spectra can be understood quite simply\cite{chen} in terms of the lowest 
band of multiplets of $N_e$ electrons weakly coupled to the hole.
There is clear evidence for bound states of the hole to one or more 
Laughlin\cite{laughlin} quasielectrons.
For $d<\lambda$ there has been no convincing interpretation of the low 
lying states, although Apalkov et al.\cite{apalkov} suggested an 
explanation in terms of ``dressed'' $X^0$ excitons.

At $d=0$ there are two types of states which contain excitons, viz.
multiplicative states containing $N_e-1$ electrons and one $X^0$, and
non-multiplicative states containing $N_e-2$ electrons and one $X^-$.
The multiplicative states are particularly simple; their energies are 
simply the energies of $N_e-1$ interacting electrons less the binding
energy $\varepsilon_0$ of an $X^0$.
The non-multiplicative states are an example of a two-component plasma 
and can be understood in our generalized CF picture.

For $N_e=7$, the $6e^-+X^0$ and $5e^-+X^-$ states can be found in the 
$8e+2h$ spectra shown in Fig.~\ref{fig2}, where they correspond to the 
$6e^-+2X^0$ and $5e^-+X^-+X^0$ multiplicative states marked with open 
symbols.
We have shown that the predictions of our model work very well for 
this system.
In particular, it is clear from Fig.\ref{fig2}ab that while the $7e+1h$ 
ground state at $2S=9$ is the (multiplicative) incompressible $\nu=2/3$ 
state of six electrons, the low lying states at $L=1$, 2, and 3 all 
contain an $X^-$ and thus their nature is very different.

Similarly, at $2S=15$, the pseudopotential calculation for the $5e^-+X^-$ 
grouping (as in Fig.\ref{fig2}bdf) as well as the CF prediction for
$m_{e^-e^-}=3$ and $m_{e^-X^-}=2$ undoubtly preclude the interpretation 
of the low energy band at $L=1$, 2, 3, and 4 (see figures in 
Refs.~\cite{chen,apalkov}) in terms of a ``dressed'' exciton in favor 
of the $5e^-+1X^-$ configuration.
In the CF picture of those states, one electron binds to the $X^0$ 
forming an $X^-$ and leaving behind a quasihole (QH-$e^-$) in the Laughlin 
$\nu=1/3$ state.
The $X^-$ (with $l^*_{X^-}=3/2$) and the QH-$e^-$ (with $l^*_{e^-}=5/2$)
have opposite charges and attract one another; what results in their
excitonic dispersion.
We have checked that the present interpretation remains valid at 
inter-layer separations $d$ up the order of $\lambda$, when $X^-$'s 
unbind (detailed analysis of spatially separated systems will be 
presented elsewhere).

\section{Photoluminescence}

A single $X^-$ cannot emit a photon by $e-h$ recombination and leave
behind a free electron.
In the simplest terms, this is because the luminescence operator 
conserves total angular momentum, and an $X^-$ has $l_{X^-}=S-1$,
while the electron has $l_{e^-}=S$.
For separated electron and hole planes, the hidden symmetry theorem
does not hold, and it is possible to have weak luminescence from an
$X^-$ interacting with other charged particles.
However, the luminescence intensity is much weaker than the fundamental
luminescence line due to a neutral $X^0$.
The existence of free $X_k^-$ complexes appears to act as a trap that
inhibits a strong luminescence intensity from $X^0$'s.
Observation of a strong $X^-$ luminescence signal seems likely to be
associated with excitons bound to an impurity and/or mixing of higher 
Landau levels.
This might break the selection rule that forbids luminescence for a 
free $X^-$.

\section{Speculation}

The generalized CF picture will be of value if it can make predictions
for systems which at the moment are too large to evaluate numerically.
An example that we have not been able to study numerically is that of
$N_e=14$ and $N_h=5$.
The configuration with the largest binding energy is (viii) $4e^-+5X^-$,
but the configuration (ix) $5e^-+3X^-+X_2^-$ is only slightly smaller
in binding energy.
Which of these configurations has the lowest energy at a given value 
of $2S$ will depend on both the binding energy and the interparticle
interactions.
For $2S=36$, we can choose $m_{e^-e^-}=m_{X^-X^-}=5$ and $m_{e^-X^-}=
m_{e^-X_2^-}=m_{X^-X_2^-}=4$.
This choice satisfies all the requirements imposed by the Pauli 
principle and by the hard cores of the different pseudopotentials.
From Eq.~(\ref{eq2}), we find $2S_{e^-}^*=2S_{X^-}^*=2S_{X_2^-}^*=4$
so that $2l_{e^-}^*=4$, $2l_{X^-}^*=2$, and $2l_{X_2^-}^*=0$.
This leads to an $L=0$ state of configuration (ix).
If it is lower in energy than the lowest state of configuration (viii),
it is very probably a Laughlin incompressible state.
For configuration (viii), we find that there is a quasihole in the 
electron shell of angular momentum $l_{e^-}^*=2$ and a pair of 
quasiparticles in the $X^-$ shell of angular momentum $l_{X^-}^*=2$.
This gives $L_{e^-}=2$, $L_{X^-}=1$, 3, and thus $L=1$, $2^2$, $3^2$, 
and 4.
It seems likely that the quasiparticle energy in configuration (viii) 
more than compensates its slightly higher binding energy and that 
configuration (ix) is an incompressible quantum fluid state.
It is unlikely that one will be able to diagonalize the nineteen 
particle electron-hole system at $2S=36$, but the nine particle systems 
($4e^-+5X^-$ and $5e^-+3X^-+X_2^-$) might be possible.

\section{Summary}

Charged excitons and excitonic complexes play an important role in 
determining the low energy spectra of electron-hole systems in a strong 
magnetic field.
We have introduced general Laughlin type correlations into the 
wavefunctions, and proposed a generalized CF picture to elucidate 
the angular momentum multiplets forming the lowest energy bands for
different charge configurations occurring in the electron-hole system.
We have found Laughlin incompressible fluid states of multi-component 
plasmas at particular values of the magnetic field, and the lowest 
bands of multiplets for various charge configurations at any value 
of the magnetic field.
It is noteworthy that the fictitious Chern--Simons fluxes and charges 
of different types or colors are needed in the generalized CF model.
This strongly suggests that the effective magnetic field seen by the 
CF's does not physically exist and that the CF picture should be 
regarded as a mathematical convenience rather than physical reality.

We thank P. Hawrylak and M. Potemski for helpful discussions.
AW and JJQ acknowledge partial support from the Materials Research 
Program of Basic Energy Sciences, US Department of Energy.
KSY acknowledges support from the Korea Research Foundation (Project
No. 1998-001-D00305).


\begin{references}

\bibitem{shields}
A. J. Shields, M. Pepper, M. Y. Simmons, and D. A. Ritchie,
   Phys. Rev. B {\bf52}, 7841 (1995);
G. Finkelstein, H. Shtrikman, and I. Bar-Joseph,
   Phys. Rev. B {\bf53} 1709, (1996).

\bibitem{wojs1} 
A. W\'ojs and P. Hawrylak, 
   Phys. Rev. B {\bf51} 10~880, (1995).

\bibitem{palacios}
J. J. Palacios, D. Yoshioka, and A. H. MacDonald,
   Phys. Rev. B {\bf54}, 2296 (1996).

\bibitem{wojs2} 
A. W\'ojs, P. Hawrylak, and J. J. Quinn, 
   Physica B {\bf256--258}, 490 (1998);
   Phys. Rev. Lett. (submitted, cond-mat/9810082);
A. W\'ojs, I. Szlufarska, K.-S. Yi, and J. J. Quinn, 
   Phys. Rev. Lett. (submitted, cond-mat/9904395).

\bibitem{kheng}
K. Kheng, R. T. Cox, Y. Merle d'Aubigne, F. Bassani, K. Saminadayar, and 
S. Tatarenko,
   Phys. Rev. Lett. {\bf71}, 1752 (1993);
H. Buhmann, L. Mansouri, J. Wang, P. H. Beton, N. Mori, M. Heini, and 
M. Potemski,
   Phys. Rev. B {\bf51}, 7969 (1995).

\bibitem{lerner} 
I. V. Lerner and Yu. E. Lozovik,
   Sov. Phys. JETP {\bf53}, 763 (1981);
A. H. MacDonald and E. H. Rezayi,
   Phys. Rev. B {\bf42}, 3224 (1990).

\bibitem{haldane1} 
F. D. M. Haldane, 
   Phys. Rev. Lett. {\bf51}, 605 (1983).

\bibitem{wu} 
T. T. Wu and C. N. Yang,
   Nucl. Phys. B {\bf107}, 365 (1976).

\bibitem{macdonald} 
A. H. MacDonald, E. H. Rezayi, and D. Keller,
   Phys. Rev. Lett. {\bf68}, 1939 (1992).

\bibitem{chen} 
X. M. Chen and J. J. Quinn, 
   Phys. Rev. B {\bf50}, 2354 (1994);
   ibid. {\bf51}, 5578 (1995).

\bibitem{rashba} 
E. I. Rashba and M. E. Portnoi,
   Phys. Rev. Lett. {\bf70}, 3315 (1993).

\bibitem{apalkov} 
V. M. Apalkov, F. G. Pikus, and E. I. Rashba,
   Phys. Rev. B {\bf52}, 6111 (1995).

\bibitem{halperin} 
B. I. Halperin, 
   Helv. Phys. Acta {\bf56}, 75 (1983).

\bibitem{laughlin} 
R. B. Laughlin, 
   Phys. Rev. Lett. {\bf50}, 1395 (1983).

\bibitem{jain}
J. K. Jain, 
   Phys. Rev. Lett. {\bf63}, 199 (1989).

\bibitem{haldane2} 
F. D. M. Haldane and E. H. Rezayi,
   Phys. Rev. Lett. {\bf60}, 956 (1988).

\bibitem{wojs3}
A. W\'ojs and J. J. Quinn, 
   Solid State Commun. {\bf108}, 493 (1998);
   ibid. {\bf110}, 45 (1999);
   Phys. Rev. B (submitted, cond-mat/9903145).

\bibitem{sitko} 
P. Sitko, S. N. Yi, K.-S. Yi, and J. J. Quinn,
   Phys. Rev. Lett. {\bf76}, 3396 (1996).

\bibitem{shalit} 
A. de Shalit and I. Talmi, 
   {\em Nuclear Shell Theory}, 
   Academic Press, New York and London 1963.

\end{references}
\end{document}